\documentclass[utf8]{frontiersSCNS} 

\usepackage{url,hyperref,lineno,microtype,subcaption}
\usepackage[onehalfspacing]{setspace}

\newcommand{\change}[1]{\textcolor{black}{#1}}

\def\keyFont{\fontsize{8}{11}\helveticabold }
\def\firstAuthorLast{Pouke {et~al.}} 
\def\Authors{Matti Pouke\,$^{*}$, Evan G. Center, Alexis P. Chambers, Sakaria Pouke, Timo Ojala and Steven M. LaValle}
\def\Address{$^{1}$Center for Ubiquitous Computing, Faculty of Information Technology and Electrical Engineering, University of Oulu, Finland }
\def\corrAuthor{Matti Pouke}

\def\corrEmail{matti.pouke@oulu.fi}

\begin{document}
\onecolumn
\firstpage{1}


\title[Body Scaling and Physics Plausibility]{The Body Scaling Effect and its Impact on Physics Plausibility} 

\author[\firstAuthorLast ]{\Authors} 
\address{} 
\correspondance{} 

\extraAuth{}

\maketitle

\begin{abstract}

\section{}
In this study we investigated the effect of body ownership illusion-based body scaling on physics plausibility in Virtual Reality (VR). Our interest was in examining whether body ownership illusion-based body scaling could affect the plausibility of rigid body dynamics similarly to altering VR users scale by manipulating their virtual interpupillary distance and viewpoint height. The procedure involved the conceptual replication of two previous studies. We investigated physics plausibility with 40 participants under two conditions. In our synchronous condition, we used visuo-tactile stimuli to elicit a body ownership illusion of inhabiting an invisible doll-sized body on participants reclining on an exam table. Our asynchronous condition was otherwise similar, but the visuo-tactile stimuli were provided asynchronously to prevent the onset of the body ownership illusion. We were interested in whether \change{the} correct approximation of physics (\textit{true physics}) or physics that are incorrect and appearing as if the environment is five times larger instead (\textit{movie physics}) appear more realistic to participants as a function of body scale. We found that \textit{movie physics} did appear more realistic to participants under the body ownership illusion condition. However, our hypothesis that \textit{true physics} would appear more realistic in the asynchronous condition was unsupported. Our exploratory analyses revealed that \textit{movie physics} were perceived as plausible under both conditions. Moreover, we were not able to replicate previous findings from literature concerning object size estimations while inhabiting a small invisible body. However, we found a significant opposite effect regarding size estimations; the object sizes were on average underestimated during the synchronous visuo-tactile condition when compared to the asynchronous condition. We discuss these unexpected findings and the potential reasons for the results, and suggest avenues for future research.

\tiny
 \keyFont{ \section{Keywords:} Virtual Reality, Perception, Scaling, Embodiment, Human Factors} 
\end{abstract}


\section{Introduction} 
Understanding size and distance perception in virtual reality (VR) has been seen as a relevant research topic, not only because of scientific interest, but also since incorrect estimations inside virtual environments (VEs) might lead to unwanted effects in various VR-based training, design and visualization applications \citep{renner2013perception}. It is a known phenomenon that familiar size cues affect the perception of sizes and distances in both real life and VR. As we see familiar objects, their known size acts as a reference in which we can compare less known features of the environment. The "body scaling effect" refers to the phenomenon of our own body acting as a familiar size cue; our own limbs act as a perceptual ruler in which we base the proportions of our surroundings and nearby objects (eg. \citep{linkenauger_illusory_2010, ogawa2017distortion}). The body ownership illusion refers to an artificial or virtual body appearing as one's own, similarly to the rubber hand illusion \citep{botvinick1998rubber}. This illusion has been used to manipulate perception of sizes and distances by making participants embody virtual bodies of different sizes in VR; as the size of the virtual body changes, the body scaling effect causes perceived sizes and distances to scale into the opposite direction \citep{van2011being,van2014body,van2016illusions,banakou2013illusory,weber2020body}. Based on their studies, van der Hoort and Ehrsson explain that the body-scaling effect is not dependent on just the visual appearance of our body, but also purely proprioceptive information manipulated by the body ownership illusion which can cause the perceptual effects of scaling sizes and distances even when visual information of the body is lacking \citep{van2014body,  van2016illusions}. In their latter experiment, this was demonstrated by eliciting a body-ownership illusion of an invisible body using a stereoscopic camera system and visuo-tactile stimuli, and examining the illusion's effect on size estimations. 

Another way to significantly affect our perception of sizes and distances is by artificially manipulating our interpupillary distance (IPD), for example, by using virtual reality \citep{kim2017dwarf}. A room modeled at human scale in virtual reality essentially appears 10 times larger when we reduce our IPD by a factor of ten by manipulating the distance of the virtual cameras acting as our eyes in VR \citep{pouke2020plausibility}. Manipulating our perception of sizes and distances also causes interesting effects regarding the perception of physical phenomena, such as rigid body dynamics. For example, an object dropped from a height of \change{1.5} m reaches the ground in \change{0.55} s whereas an object dropped from 15 cm reaches the ground in only \change{0.17} s. Although this behavior seems natural in everyday life, the phenomenon appears greatly unnatural when viewed at abnormal scales in VR, even when we are fully aware of \change{being scaled} \citep{pouke2020plausibility, pouke2021plausibility}. Difficulties in perceiving physics at abnormal scales have been shown to increase difficulties in, for example, robotic teleoperation at micro- and nanoscopic scales \citep{millet2008improving, sitti2007microscale, zhou2000virtual}.




\subsection{Perception of sizes and distances}
As discussed above, the size cues of an environment affect our perception of sizes and distances. Previous research has suggested that egocentric distances are generally underestimated in VR due to a multitude of factors. For example, richer environmental cues and realistically modeled environments generally seem to improve the accuracy of distance judgements \citep{renner2013perception} and sensitivity to height perception \citep{deng2019floating}.  \cite{langbehn2016scale} studied mismatching size cues in virtual environments and found that participants generally relied on their own bodies when judging the correct scale. An exception, however, was the presence of multiple virtual characters, when the scale of the virtual characters was perceived as the correct one.


Previous research also suggests that our own action capabilities can affect our perception. For example, wearing a heavy backpack makes a hill appear steeper since the heavy backpack alters our perceived affordances \citep{bhalla1999visual}. However, these studies have been criticized, claiming instead that the identified affordance-based perceptual effects have been due to demand characteristics and not to actual changes in perception (eg. \citep{hutchison2006does}).

The body-scaling effect refers to the phenomenon of our own body, real or virtual, affecting the perception of sizes and distances. \cite{linkenauger_illusory_2010} manipulated the retinal size of objects using magnifying and "minifying" goggles and found that placing a hand next to these objects appeared to scale the objects back towards their normal size. \cite{ogawa2017distortion} found that not only can our hand size affect the perceived size of objects, but familiar sized objects can also affect the perceived size of hands in VR. In a follow-up study, \cite{ogawa2019virtual} reported that the strength of the body-scaling effect can be altered by the visual realism of the virtual hand.

The body ownership illusion refers to the sensation of a virtual or artificial body appearing as one's own body, similarly to the rubber hand illusion \citep{botvinick1998rubber}. In previous research, one of the most popular ways of achieving the body ownership illusion has been through the application of synchronous visuo-tactile stimuli, for example by touching the artificial body simultaneously with the corresponding location in the participant's own body. Typically, asynchronous stimulation, performing the touches out of sync, has been the control condition in these types of experiments. According to \cite{maselli2013building}, visuomotor synchrony and appearance of the virtual body can also be utilized to elicit the illusion. In addition, the body ownership illusion can even take place purely without visuo-tactile stimuli; however, including visuo-tactile stimuli can be helpful for eliciting the illusion when other properties of the illusion (such as first person perspective or the visual appearance of the virtual body) are violated \change{( \citealt{maselli2013building})}.

The effect of a full body ownership illusion \citep{slater2009inducing} for size and distance perception has been the focus of multiple studies. Van der Hoort and Ehrsson embodied participants as dolls and giants using head-mounted displays (HMDs), stereoscopic cameras, and synchronous visuo-tactile stimuli, and found that body ownership illusions significantly affected the perceived sizes of nearby objects. Later, Banakou et al. used VR to embody participants in a child's body and found it not only affecting the perceived size of objects, but also participants' associated personality traits \citep{banakou2013illusory}. Serino et al. found that embodying bodies of extreme sizes affected participants' judgements of the properties of their own bodies \citep{serino2020gulliver}.

According to later studies by van der Hoort and Ehrsson, the body scaling effect caused by the body ownership illusion exists even when the body is not visible, suggesting that the effect is more related to proprioception than vision \citep{van2014body,van2016illusions}. In the latter study, which is conceptually replicated in this paper, illusions of inhabiting small and large invisible bodies were elicited among participants. The participants were lying down on a bed wearing HMDs, which streamed a stereoscopic image from two cameras placed on a bed in a laboratory. The participants' legs were then stroked while simultaneously moving a brush in front of the camera either 60-80 cm away (small body) or 300-400 cm away (large body). The experimenters then found an inverse relationship between the body size and estimated object sizes during synchronous stimuli, whereas this effect did not exist in the control condition in which the touch and brush movement were asynchronous. The work of \change{\cite{banakou2013illusory}} found a similar perceptual effect when the body ownership illusion was controlled by visuomotor synchrony. The work of \change{\cite{weber2020body}} found object sizes were estimated differently across three different body size conditions; however, they did not find differences between synchronous and asynchronous stimulus conditions. This is in line with the findings of Maselli and Slater that \change{a} synchronous visuo-tactile stimulus is not necessary for the body ownership illusion in case\change{s where} a visible body is experienced from a first-person perspective \citep{maselli2013building}. Moreover, according to their findings, even asynchronous stimuli can be perceived as real when other properties for body ownership illusion are taking place. 

We are not aware of any study that would have investigated this perceptual effect with an invisible body in VR. Moreover, we are, as of now, unaware of any studies that would have investigated the effect of body scaling on the perception of physics\change{,} as the studies of \cite{pouke2020plausibility, pouke2021plausibility} utilized IPD-based scaling and not body scaling.



\subsection{Perception of physics}
In terms of \change{perceiving} rigid body dynamics, different scales appear similar to being under the influence of non-earth gravity. When perceiving object velocities and accelerations, being ten times smaller appears essentially the same as if gravity had turned tenfold, while scales further away than one order of magnitude introduce even more peculiarities (see for example \citep{zhou2000virtual}). However, it appears that humans have the tendency to instinctively expect rigid body dynamics to behave similarly to human-scale under normal gravity conditions.  \cite{mcintyre2001does} studied this phenomenon by experimenting with astronauts performing in zero gravity and found that their ability to catch vertically moving balls was less accurate in comparison to normal gravity conditions. Human capacities for intercepting moving objects under various directions and accelerations was further analyzed by \cite{senot2005anticipating} in a VR experiment. They also found evidence of humans being more capable of intercepting objects behaving as if under normal gravity, even if best success rate was achieved in intercepting objects under constant velocity. \cite{jorges2017gravity} argued that normal earth gravity is a strong Bayesian prior in human perceptual processes, and therefore any contradicting evidence is perceived as false. This bias then leads to generally poor human performance in non-normal gravity conditions and renders adaptation difficult. 

\begin{figure}[h]
 \centering 
 \includegraphics[width=0.8\columnwidth]{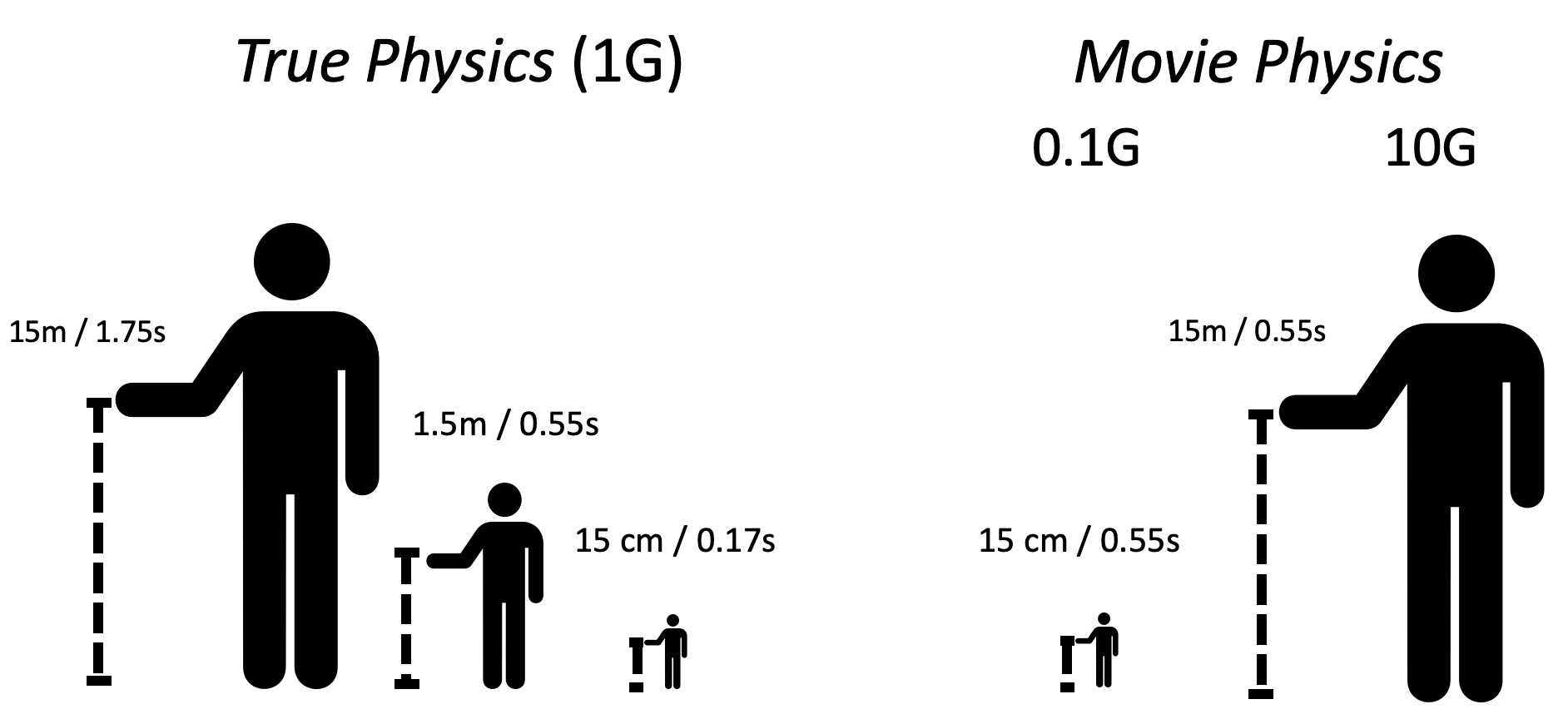}
 \caption{\change{A figure depicting conditions reported by \cite{pouke2021plausibility} to illustrate the relationship between gravity and scale among various pairs. In the true physics condition, gravity behaved realistically at each scale (left). In the movie physics conditions, gravity was manipulated so that free-fall speeds of objects (and rigid body dynamics in general) matched that of human height, regardless of scale (right).}}
 \label{conceptual}
\end{figure}

In the field of VR research, \textit{fidelity} is referred to as the extent to which the VR system faithfully simulates the real world. \textit{Plausibility illusion}, on the other hand, refers to an illusion of realism experienced subjectively by the user \citep{skarbez2017survey}. Fidelity does not necessarily lead to plausibility, as plausibility depends on the expectations of the user instead of physical realism, and can be affected by priming as well as the context of the virtual environment (VE). \cite{skarbez2017psychophysical} suggested a concept called \textit{coherence}, which, instead of fidelity, refers to the properties of the VE that affect the onset of plausibility illusion. In our previous research \citep{pouke2020plausibility}, we have studied the plausibility of physics models at abnormal scales by virtually scaling participants both ten times smaller and ten times larger by manipulating their IPD, viewpoint height and motion controller interaction distance. Participants dropped and threw objects and their plausibility was estimated under two physics conditions: realistic approximation of physics (a higher fidelity model dubbed \textit{true physics}) and an inaccurate model that functioned as if the world had changed in size and participants remained at normal scale (a lower fidelity model dubbed \textit{movie physics} \change{equaling 0.1G at small scale and 10G at large scale, see Fig \ref{conceptual}}). As a result of scaling, \textit{true physics} appeared as fast object accelerations and short throwing distances in the small-scale study, and slow accelerations and large throwing distances in the large-scale study. We queried plausibility using two forced-choice questions. The first question queried which one of the models the participants considered matching actual reality, whereas the second question asked which one of the models the participants considered matching their expectations better. \textit{Movie physics} was chosen as the realistic model by roughly 70\% of participants. Interestingly, however, in the small scale study, roughly 90\% of participants considered \textit{movie physics} to better match their expectations, whereas in the large scale study neither \textit{movie physics} nor \textit{true physics} were chosen significantly more often as the model matching expectations better. This could mean that in the small-scale study, there were participants who considered \textit{true physics} surprising even if they considered it to be more realistic. In the large-scale study, however, the opposite appeared to happen; some participants found \textit{true physics} \change{to match} their expectations better even if they ultimately considered \textit{movie physics} as the more realistic model. In any case, it can be roughly summarized that in these studies, \change{participants perceived high fidelity settings as having low coherence and low fidelity settings as having high coherence}. \cite{pouke2020plausibility, pouke2021plausibility}

\change{In this paper, we present our results of investigating whether body ownership illusion-based body scaling can alter the perception of physics similarly to IPD based scaling as reported by \cite{pouke2020plausibility}. This work extends the previous work of \cite{pouke2020plausibility,pouke2021plausibility}, generating new knowledge on plausibility of physics perception by investigating previously unexplored factors. In addition, this work acts as a conceptual replication study for the work of \cite{van2016illusions}, investigating whether the invisible body illusion replicates in VR.}











\section{Materials and Methods}
The motivation for this study was to follow up on our two previous studies to address unexplored factors regarding perception of physics in scaling. Our objective was to investigate whether body scaling caused by body ownership illusion could affect the perceived naturalness of object motions similarly to IPD-based scaling. 


We designed a VR experiment based on the invisible-body illusion experiment reported by \cite{van2016illusions} as well as the scaled-down physics perception experiment by \cite{pouke2020plausibility}. We adopted our physics conditions from the work of \cite{pouke2020plausibility}, with \textit{true physics} \change{serving as} the realistic approximation of physics, and \textit{movie physics} representing physics that \change{behaved as though} the environment had been scaled up instead ('movie' referring to physical behavior typical in Hollywood movies depicting scaled-down characters and their physical surroundings). We chose to utilize the visuo-tactile induced invisible body illusion in the vein of \cite{van2016illusions} as a way to investigate the body ownership illusion and body scaling while avoiding possible confounds related to having a visible body. As in the original experiment, we controlled the invisible body illusion using either synchronous or asynchronous visuo-tactile stimuli. Using a visible body, however, it would have become possible that the participants would have experienced the body ownership illusion also in the asynchronous condition due to merely seeing the virtual body (for example, \citep{maselli2013building, weber2020body}).




\subsection{Conditions and hypotheses}
Our conditions followed the experimental protocol of \cite{van2016illusions}, providing either synchronous or asynchronous visuo-tactile stimuli to control the onset of an invisible body illusion of possessing a small body, and an object size estimation task to assess the strength of the body scaling effect. To investigate the effect of body scaling on physics perception, we added a physics plausibility estimation task similar to \cite{pouke2020plausibility} into both syncronous and asynchronous conditions before each object size estimation task. To keep the experimental conditions \change{from} deviating too much from the protocol of \cite{van2016illusions}, we asked the participants to observe a virtual character \change{performing} the interaction task instead of asking the participants to interact with the objects themselves, similarly to our previous studies. Also, differing from our previous studies, we utilized only one forced-choice question, instead of two, to query for plausibility: more specifically, we asked the participants to state the model that \change{better matched their expectations}. This choice was made because we wanted the participants to specifically report on their first impressions, instead of stopping to consider which one of the models should be real. There would have been little point \change{asking about realism}, given that the physics estimation task in both synchronous and asynchronous conditions was visually identical. Other differences between this study and the aforementioned studies can be summarized as follows: Unlike  \cite{van2016illusions}, our experiment utilized VR instead of physical cameras and objects. In addition, we only considered a small-body condition rather than both small-body and large-body conditions \change{to keep the total number of counterbalanced conditions manageable (please see the details about our preregistration at the end of this subsection)}. Finally, differing from \cite{pouke2020plausibility}, our scale of interest was only 5 times smaller than human scale since a 10 times smaller invisible body made the visuo-tactile stimuli too difficult to observe without manipulating IPD. This made our invisible body roughly between 30-40 cm in size, depending on the height of the participant, which was roughly the same size as the smallest doll body used by \cite{van2011being}.



Our hypotheses were as follows:

H1: After synchronous visuo-tactile stimuli, subjects will consider \textit{movie physics} to appear more real.

H2: After asynchronous visuo-tactile stimuli, subjects will consider \textit{true physics} to appear more real.

These predictions are driven by the idea that participants should use their perceived body scale to predict their environment's physics. Thus, when they are instilled with the illusion of inhabiting a small body, they should estimate \textit{movie physics} to be more plausible, whereas when they are left with their normal body scale, they should estimate \textit{true physics} to be more plausible. In addition to physics plausibility observations, we also collected object size estimations for estimating the effect of body scaling, as well as presence and background data for exploratory purposes. We preregistered our procedure and analysis methods at osf.io \footnote{\url{https://osf.io/4gjv3/?view_only=b20e3e383852456eafb1225e34a85617}}. It should be noted that our preregistration also includes two additional hypotheses that are related to two additional conditions regarding physics perception at abnormal scales. \change{In response to reviewer feedback and efforts to improve clarity}, however, the results and other details regarding these two hypotheses and are reported in a separate manuscript instead (Pouke et al., in preparation).

\subsection{Participants}
The experiment was a within-subject design with 40 participants (20 females and 20 males) naive to the purposes of the experiment. The participants gave written informed consent in accordance with the local ERB. The sample size was based on our previous studies on perception of physics plausibility (\cite{pouke2020plausibility},  \cite{pouke2021plausibility}). This sample is also twice as large as what was reported as necessary to detect the perceptual effects of body scaling by \cite{van2016illusions}. We had to replace 4 participants due to significant differences in experimental conditions due to researcher and software errors, bringing the total number of participants we ran to 44. \change{The participants' ages ranged from 18 to 46, the average age being 27,5. The participants self-reported their VR and video game experience using a scale from from one (no experience) to seven (daily use). Using this quantification, their average VR and video game experience was reported as 1.8 and 4.15, respectively.} For Covid-19 related precautions, all equipment and surfaces were disinfected using alcohol wipes. The Cleanbox device was used to disinfect the HMD between participants \footnote{https://cleanboxtech.com/}. Researchers were wearing masks throughout the experiment. Masks and hand sanitizer \change{were} also available for participants.

\subsection{Experimental apparatus and protocol}
\begin{figure}[tb]
 \centering 
 \includegraphics[width=\columnwidth]{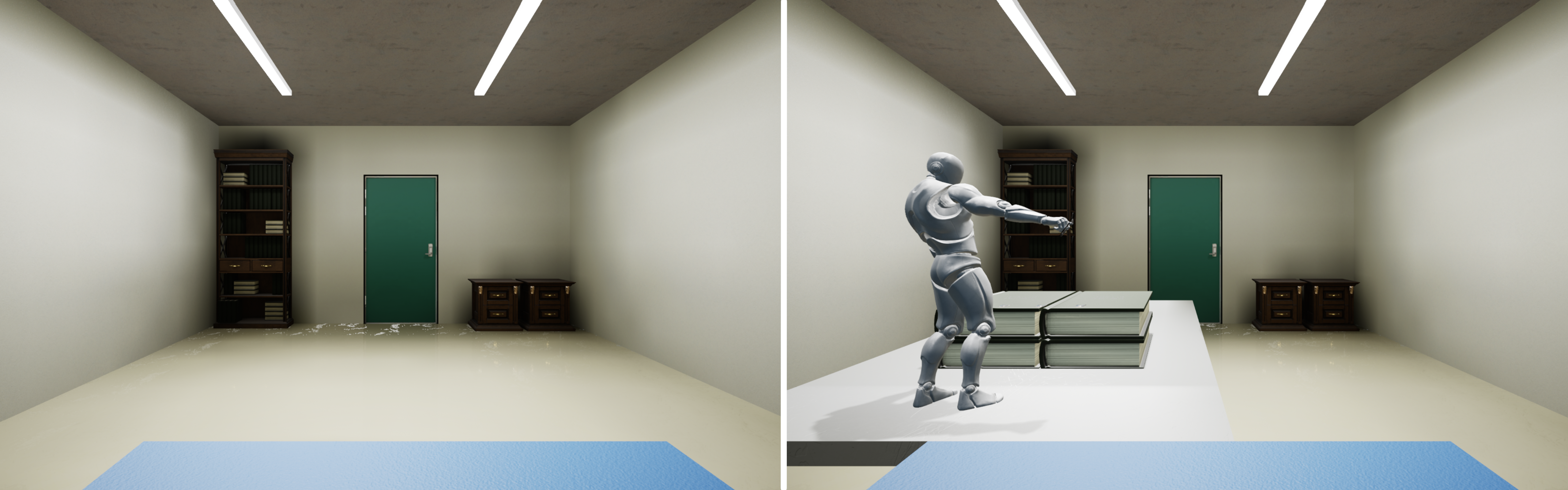}
 \caption{Virtual environment as seen by participants: during stimulation and cube estimation phases (left), during physics estimation task (right).}
 \label{room}
\end{figure}

Using Unreal Engine 4, we prepared an experimental application \change{designed} to loosely mimic the laboratory conditions reported in the work of \cite{van2016illusions}. The virtual environment depicted a simple room with a bed on one side, and a door, bookshelf and a nightstand at the opposite side of the room (see Fig. \ref{room} top left). The virtual camera was placed on a virtual exam table looking towards the door. The height of the camera was set so that it matched the reclined angle in which the participants were sitting, but at 0.2 scale.

As stated in previous sections, the physics conditions were demonstrated by an animated robot. We chose to use a robot instead of a human-looking avatar to help maintain more consistent size cues for the environment; we believed an animated robot could be perceived as something akin to an animatronic puppet. In addition, we considered using the default game engine assets where possible for the benefit of the replicability of this study. 
The animations of the robot were captured using Vive Trackers and the Unreal Engine Vive MoCap plugin. A researcher wearing the Vive Trackers performed a sequence of picking up and dropping three tabs and throwing two. The MoCap plugin was used to target this sequence into the Unreal Engine default mannequin that acted as our robot.  The same animation sequence was used for all subjects and all conditions to prevent confounds. The rigid body dynamics of the pop tabs was simulated using the built-in physics engine in Unreal Engine. Similarly to \cite{pouke2021plausibility}, physics were controlled by manipulating gravity; the prototype switched between physics conditions according to the experiment protocol by loading identical-looking levels with different World Gravity settings. For \textit{true physics}, gravity was set at 1g whereas for \textit{movie physics} it was set at 0.2g, essentially scaling rigid body dynamics of objects as if the doll-sized robot was human-sized. Using animation Notify States, the robot was programmed to pick up and release the pop tabs at specific animation frames utilizing the default 'grab' and 'throw' events that are preprogrammed inside the Unreal Engine VR "Motion Controller" template for interactive picking up and throwing in VR. This means that the pop tab objects were attached to the robot arm after a 'grab' event and detached during 'throw' events, after which the physics engine simulated their trajectory according to the object velocity at time of release. This made the simulation of throwing and dropping the tabs similar to how object throwing and dropping was simulated in our previous studies, with the only exception being that the physical dropping and throwing motions were prerecorded instead of performed by each participant individually. As in our previous studies, we also assumed the mass of the pop tab light enough not to affect the physical arm motions due to inertia or lack of muscle strength \citep{cross2004physics}. 

For VR hardware, we used a Valve Index HMD and controllers.

\subsubsection{Stimuli}
During the synchronous condition, the participant received synchronous visuo-tactile stimuli while reclined on an exam table. The stimuli were provided in an attempt to provide an invisible-body illusion with a body scaled down at 20\% of original size. The stimuli were provided using a motion controller and a tennis ball. Previous research (eg. \citealt{van2016illusions}) has stated that visuo-tactile stimuli administered to the feet and lower legs are useful for eliciting a full-body illusion and we also followed this procedure. While we touched the participants' lower legs and feet, a yellow ball was moving within the virtual environment. However, the transformations of the virtual ball were scaled down and offset so that while the participants legs were touched with the physical tennis ball, the virtual ball appeared to be only tens of centimeters away from the participant's head. The offset required for the transformations of the virtual tennis ball was derived using a avatar scaled down to 0.2 from human size reclined on the virtual exam table; this avatar was obviously deleted during the actual experiment. In addition, the location of the ball was calibrated prior each participant to accommodate for participants of different heights. The asynchronous condition was identical to the synchronous with the exception that the visuo-tactile stimulus was asynchronous, moving the controller and the physical tennis ball independently of each other, to break the correspondence between visual and tactile senses. \change{In both conditions, we utilized a combination of sweeps and taps at different locations of the participants' feet and lower legs while making an effort to keep the movements of the ball somewhat random and unpredictable (inspired by the procedure described by \citealt{slater2008towards}). In the asynchronous condition, the experimenter was holding the controller in one hand and the ball in another, so that he, or she could produce both temporal delays as well as spatial asynchronicity between the visual and physical stimuli in a randomized manner while otherwise mimicking the characteristics of the taps and sweeps introduced in the synchronous condition (please see the video link appearing later in this section).}

After 90 seconds of stimulus presentation, the virtual ball was hidden, and a table, a scaled-down robot, a stack of books and a set of soda can tabs appeared (See Fig. \ref{room} right). Each participant then experienced animation sequences of a doll-sized robot (similar \change{in} size to the simulated invisible body) handling pop tabs for the purposes of physics plausibility estimation. The robot picked up and dropped three tabs, and threw two across the book. The animation was presented twice, with the behavior of tabs following either \textit{true physics} or \textit{movie physics}, depending on the order of conditions.The order of the true and movie physics presentations, as well as the synchronous and asynchronous conditions in which they were nested, were counterbalanced among participants. For full details regarding the counterbalancing scheme, including conditions participants subsequently experienced as part of protocols for Pouke et al. (in preparation), see Appendix 1.  After viewing both types of physics, the participant was asked to perform a verbal judgement on the perceived plausibility of the tabs behavior using the question adapted from  \cite{pouke2020plausibility}: \textit{"Thinking back how the pull tabs were behaving, which matched your expectations, the first or the second time?"}.

\begin{figure}[tb]
 \centering 
 \includegraphics[width=\columnwidth]{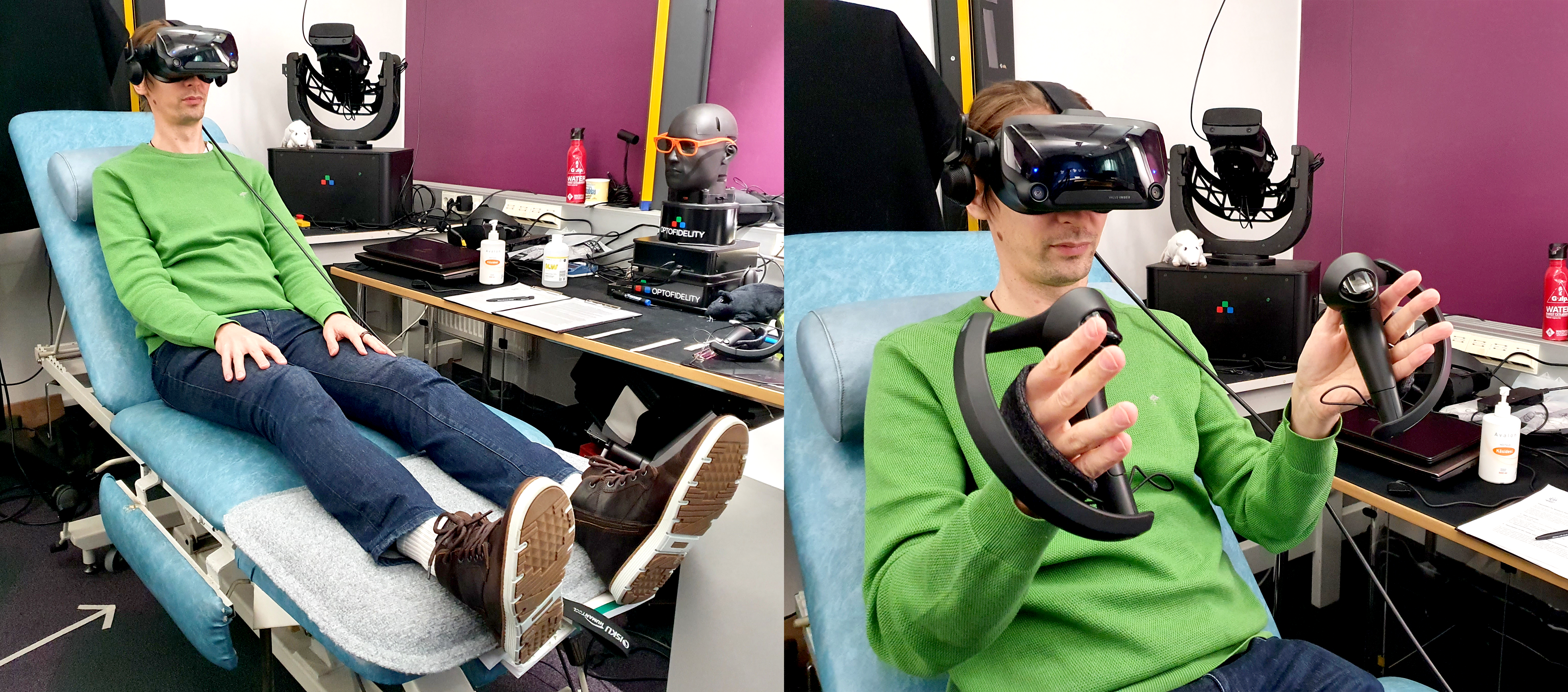}
 \caption{Participants were resting on an exam table during synchronous and asynchronous visuo-tactile conditions (left) and using VR controllers to estimate cube sizes (right).}
 \label{aku}
\end{figure}

After the verbal judgement, the physics perception related objects were again hidden, and 60 seconds of stimuli were provided. After this, the participant was given the motion controllers, and presented with three green cubes (10 cm, 20 cm and 30 cm) appearing in randomized order at a distance of 15 cm. After a cube had gone out of sight, we asked the participant to show the apparent size of the cube bimanually (see Fig. \ref{aku}, right) and to click a trigger button to save their response, which was followed by the presentation of the next cube. After estimating the size of three cubes, the participant removed the HMD and controllers and filled out an illusion strength questionnaire adapted from \cite{van2016illusions}. The procedure was then repeated for either the synchronous or asynchronous condition, depending on the counterbalancing order. \change{A video demonstrating the experimental conditions, including the administration of both synchronous and synchronous stimuli, can be viewed at: \hyperlink{https://youtu.be/IxLSxH6ZqvU}{https://youtu.be/IxLSxH6ZqvU}.}

After finishing both conditions, the participant removed the HMD and filled in a post-experiment questionnaire consisting of the Slater-Usoh-Steed presence questionnaire (SUS) \cite{slater1994depth,usoh2000using} and a background questionnaire. At the conclusion of the experiment, the participant was debriefed and compensated with a gift card worth 20 \texteuro.

\section{Results}
\subsection{Confirmatory analysis}
We obtained unexpected results regarding the effect of body scaling on the perception of physics plausibility. We predicted that \textit{movie physics} would appear more realistic in the synchronous condition and that \textit{true physics} would appear more realistic in the asynchronous condition. However, according to responses to the main question, \textit{movie physics} were selected by a majority in both conditions (synchronous 28/40, asynchronous 30/40). Following the preregistered procedure, we tested both hypotheses using the binomial test (one-tailed) against the null proportion (20/40). According the results of the tests, H1 was supported (condition A, $p = 0.008$) while H2 was unsupported (condition B, $p = 1.00$).

While the evidence in favor of H1 technically indicates that \textit{movie physics} do indeed appear more real while embodied, the lack of support for H2 raises questions whether the body scaling effect was the actual reason why \textit{movie physics} were preferred. Since rigid body dynamics under normal earth gravity \change{are} usually perceived as natural (eg. \cite{jorges2017gravity}), it seems likely that the physics perception was modulated by something other than the visuo-tactile stimuli.



\subsection{Exploratory analysis}
\change{Here we present further analyses that were not preregistered, along with Bayes factors (reported as $BF$) to accompany each measure. Unlike frequentist statistics, an advantage of the Bayesian approach is that it can estimate support for the null hypothesis in addition to support for the alternative (eg., see \citep{wagenmakers2017need}). Bayes factors provide evidence on a continuous scale, though conventionally, a value of 3 or more is considered support for the alternative hypothesis, 0.33 or less is considered support for the null hypothesis, and values between these limits signal that the data are insensitive to distinguishing between hypotheses, with values between 1 and 3 and values between 0.33 and 1 considered merely anecdotal evidence in favor of the alternative or null hypothesis, respectively \citep{jeffreys1961theory, lee2014bayesian}. We used R statistical software to estimate Bayes factors for proportion tests \citep{morey2021bayes}, Wilcoxon signed-rank tests and Spearman's $\rho$ \citep{van2020bayesian}, and binomial regression models \citep{burkner2017brms, makowski2019btr} using function-default priors, unless otherwise noted.}

We performed an additional two-tailed binomial test to investigate whether \textit{movie physics} were also preferred in the asynchronous condition. The test revealed that \textit{movie physics} were significantly preferred ($p = .002$). Therefore, \textit{movie physics} were significantly preferred across both conditions \change{(synchronous $BF = 6.18$; asynchronous $BF = 32.75$). Furthermore, if we update our prior such that the ratio of preference for \textit{true physics} in the asynchronous condition matches that of \textit{movie physics} in the synchronous condition, we find very strong support for the alternative hypothesis ($BF > 100$).} This further confirms that contrary to our \change{predictions}, \textit{movie physics} were preferred by a significantly greater number of participants regardless of condition.


\subsubsection{Estimation of cube sizes}
\begin{figure}[tb]
 \centering 
 \includegraphics[width=\columnwidth]{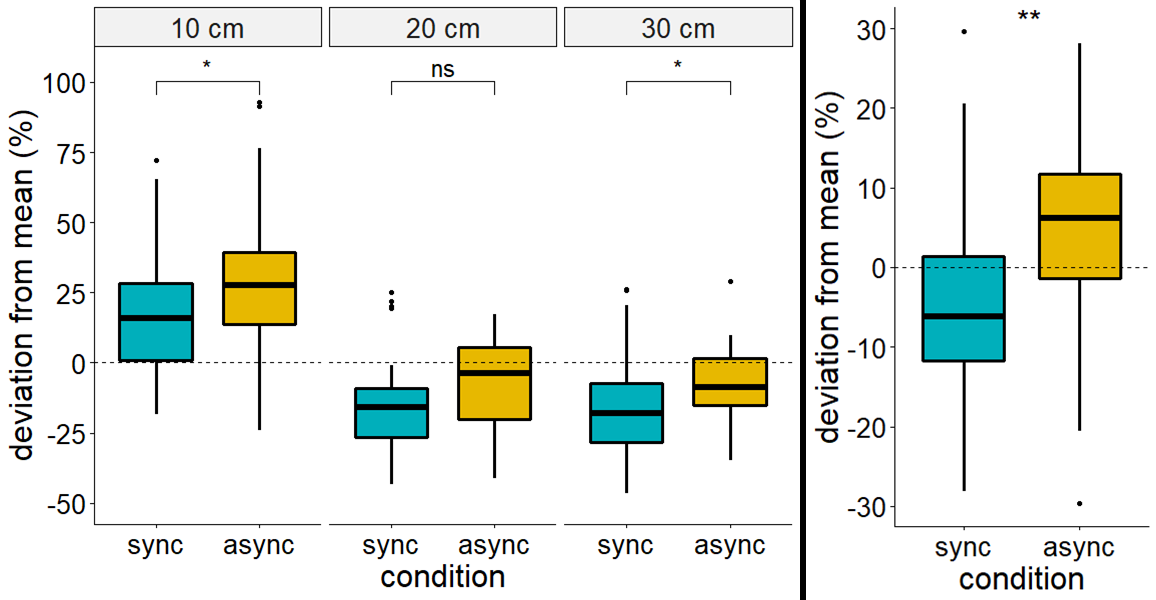}
 \caption{Boxplots for cube size estimation data. Deviations from mean across conditions per individual cubes (left) and total deviation from mean across condition (right).}
 \label{cubes}
\end{figure}

The purpose of the cube estimation task was to investigate whether the synchronous condition affected the perception of sizes similarly to \cite{van2016illusions}. To replicate their protocol, we first re-scaled all estimates to the level of the largest cube (30 cm), and then standardized the estimates to each participant's average estimate across all six conditions. This transformation allowed us to investigate deviations of estimates in percentages relative to each participant's perceptual mean, rather than in raw centimeters, thus producing the same "deviation from mean" measure touted in the original study. Negative values on this scale represent underestimations of object sizes relative to a participant's perceptual mean, whereas positive values represent overestimations. A Kolmogorov-Smirnov test on the difference of the standardized distributions indicated that the data was non-normal ($D = 0.67, p < .001$), thus we used non-parametric Wilcoxon signed-rank tests to analyze the differences between estimations. Surprisingly, we found effect opposite to the one found in \citep{van2016illusions}; the estimates for cube sizes were significantly larger in the asynchronous condition ($Z = 2.82, p = .005, r = .45, \change{BF = 5.91}$) (see Fig. \ref{cubes}), meaning that participants tended to overestimate object sizes in the condition in which the body ownership illusion should have been absent, yet underestimate when it should have been present. 

Investigating the estimations across different cube sizes, it appears the smallest cube (10 cm) was overestimated in both conditions, the overestimation being larger in the asynchronous condition. As for medium (20 cm) and large (30 cm) cubes, however, it appears that the cubes were approximated closer to their actual size in the asynchronous condition, while the sizes were underestimated during the synchronous condition. Boxplots for cube estimation data can be seen in Fig. \ref{cubes}.

The results of the size estimation tasks were thus surprising, as well. It appeared that the visuo-tactile stimulation did have a significant perceptual effect in line of \cite{van2011being, van2014body, van2016illusions}. However, some unforeseen aspect of the experiment reversed the expected outcome. 

\subsubsection{Illusion strength}
\begin{figure}[tb]
 \centering 
 \includegraphics[width=\columnwidth]{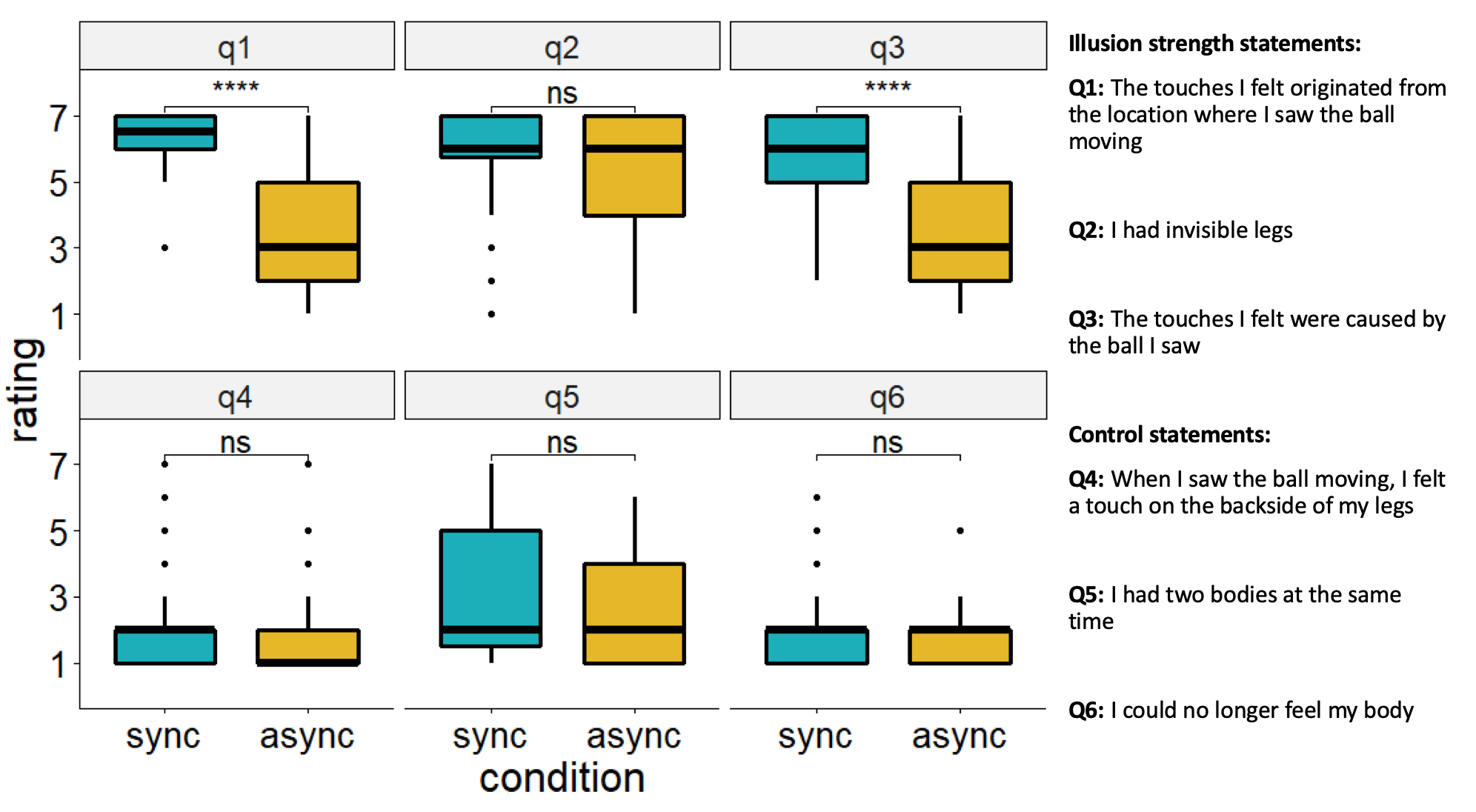}
 \caption{Boxplots of the ratings for each statement in the illusion strength questionnaire. }
 \label{emb_questionnaire}
\end{figure}

We utilized the illusion strength questionnaire from  \cite{van2016illusions} to investigate the strength of the invisible body illusion in synchronous and asynchronous conditions. Similarly to their findings, Wilcoxon signed-rank tests revealed a significant difference in statements 1 ($Z = 4.99, p < .001, r = .79, \change{BF > 100}$) and 3 ($Z = 4.53, p < .001, r = .72, \change{BF > 100}$) and no difference in control statements 4-6 \change{(all $ps > .05$; q4 $BF = 0.25$, q5 $BF = 0.18$, q6 $BF = 0.18$)}. Surprisingly, there was no significant difference in statement 2, "I had invisible legs", the median response being 6 in both conditions ($Z = 1.48, p = .14, r = .23, \change{BF = 0.42}$). Boxplots for illusion strength questionnaire data can be seen in Fig. \ref{emb_questionnaire}. The illusion strength questionnaire results are mostly in line with the results reported by \cite{van2016illusions}. This implies that the application of both synchronous and asynchronous stimuli were experienced as intended; the participants experienced being touched by the virtual ball despite the virtual ball being offset much closer to them than their real feet. In addition, this \change{illusion} was not experienced during the asynchronous condition. \change{Curiously,} however, the responses to statement 2 \change{suggest} that the participants might have felt ownership towards an invisible body regardless of stimulus condition. \change{Here an examination of the pattern of support for null hypotheses regarding illusion strength questionnaire statements may aid interpretation. Notably, the associated Bayes factors provide moderate support for the lack of a true difference between synchronous and asynchronous conditions for statements 4-6, yet only weakly support the same lack of an effect for statement 2.}

\begin{figure}[tb]
 \centering 
 \includegraphics[width=0.5\columnwidth]{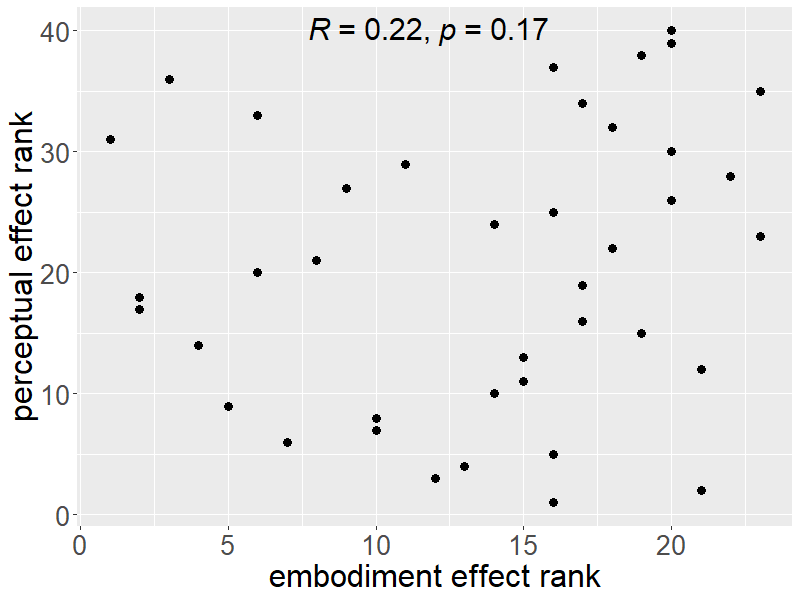}
 \caption{Correlation plot between invisible body illusion (Q1-Q3) and perceptual effect. No significant correlation was found neither when inspecting the overall correlation across all questions nor per individual questions.}
 \label{correlation}
\end{figure}

We also investigated the relationship between cube size estimations and illusion strength using the method described in \cite{van2016illusions}. Each participant's perceptual effect was quantified as the mean of their synchronous condition size estimates minus the mean of their asynchronous condition size estimates, divided by their overall estimation mean. Likewise, each participant's embodiment effect was quantified as the mean of their responses to illusion strength questionnaire items 1-3 minus the mean of their responses to items 4-6. Participants were rank ordered for each effect dimension, respectively, and the association between the effect ranks was analyzed using Spearman's rank correlation. However, unlike \cite{van2016illusions}, we found no correlation between the strength of the invisible body illusion and the perceptual effect ($p = .17, \change{BF = 0.34}$; see Fig. \ref{correlation}). This outcome \change{suggests} that the strength of the invisible body illusion \change{was not closely tied to} the strength of the perceptual effect experienced by the participants, \change{although the associated Bayes factor was not strong enough to provide more than anecdotal evidence in favor of the null hypothesis.}


\subsubsection{SUS and background}
Similarly to \cite{pouke2020plausibility, pouke2021plausibility}, we collected PI data using the extended version of the SUS questionnaire \citep{slater1994depth, usoh2000using} as well as background information, to investigate whether these factors influenced the preference of physics choices. We computed the SUS score as the sum of 6 and 7 responses per participant. The background data included age, gender as well as video game and virtual reality experience quantified as scores between 1 and 7. Similarly to our previous studies, we found none of these factors predicted the physics preferences; there were no significant associations among any of the physics choices and \change{SUS or} any of the demographics variables \change{in binomial regression models (all $ps > .05$; however, data were insensitive as all $BF$s fell between 0.3 and 3.0 relative to null models)}.

\section{Discussion}
Our main analyses provided evidence for H1. However, we found no support for H2. Our exploratory analysis found that \textit{movie physics} were perceived as more realistic across both conditions. Thus, in all likelihood, the body ownership illusion was not the primary driver behind the outcome of H1. 

\subsection{Physics judgement}
The preference for the \textit{movie physics} in the asynchronous condition came as a surprise to us. In the synchronous condition, we anticipated that the body-ownership illusion based body-scaling effect would affect the perception of physics plausibility, making \textit{movie physics} appear more real. In the asynchronous condition, however, there was no intended manipulation of size and distance perception, yet the outcome was the same as in the synchronous condition. In addition, had the inducement of the body-ownership illusion somehow failed, we would have anticipated \textit{true physics} to become the preferred physics model. 

While additional experiments will be necessary to fully account for the popularity of the incorrect physics judgements, we can speculate some of the reasons that might explain the results. An obvious question is whether the preference towards \textit{movie physics} could be explained by the fact that statement 2 of the illusion strength questionnaire, "I had invisible legs" received high scores in both conditions. If the participants experienced the body ownership illusion in both conditions, it could, in principle, explain why \textit{movie physics} were experienced as plausible in both conditions. \change{However, the associated Bayes factor for statement 2  revealed only anecdotal support for equality between the synchronous and asynchronous conditions ($BF = 0.42$), and furthermore, the perceptual effects regarding object size estimations do not seem to support this interpretation as the body scaling effect appears to have reversed.}  

Another explanation might be that we used a humanoid robot performing the object manipulation tasks. \cite{langbehn2016scale} reported that the existence of virtual characters acted as a strong size cue in their experiment, even overriding the participant's own virtual body. In our experiment, the only visible body was the one belonging to the robot. It could be that in our experiment the participants did not relate the trajectories of objects to their own (real or virtual) bodies, but to the robot's body instead, even when the robot was obviously smaller than their physical selves \change{(the robot was also surrounded by size cues indicating its size, such as a stack of books and a table)}. Therefore, it was the robot body that became the perceptual ruler to the participants, at least in relation to object motions. \change{This is similar to the findings from our previous studies \cite{pouke2021plausibility}, in the sense that the participants preferred object motions according to which the handler of the objects (controlled by themselves in our previous study) would have been at human scale, despite other size cues telling otherwise (in our previous studies, we also directly told the participants they are up/downscaled). It would appear as if participants in all three studies considered the environment scaled, and the source of object motions unscaled, when judging object motions, even if they "know" that the environment should be unscaled.}

Another potential factor is related to the animation performed by the robot which is based on motion capture (as explained in Section 3.3). Since the robot was animated using rather natural-looking human motions, it might reinforce the illusion that all motions are taking place in human scale. Perhaps, if the animation was more artificial-looking, "robot-like" or resembling the motions of a mouse-sized creature (especially regarding the speed of which the robot's limbs are moving), the naturalness of \textit{true physics} would have become more apparent. In addition, if the object motions were caused entirely by some non-humanoid entity, the plausibility of object motions might have been perceived differently.

\subsection{Object size estimation}
We also collected object-size estimation data to confirm that the body-ownership illusion we were causing was affecting the participants' perception of sizes. These results were surprising as well. Instead of the synchronous condition making objects appear larger, it appears as if they were underestimated in that condition, instead. This also warrants future studies in order to explain what caused the inverse effect.

We can look at the differences between this study and that of \cite{van2016illusions} to speculate on what caused the inverse effect. Firstly, this study took place in VR whereas the original study used a stereoscopic camera system instead. However, the body-scaling effect is known to exist in VR as well. \cite{banakou2013illusory} also found object sizes appearing larger to participants when embodying the virtual body of a 4-year old child. \cite{weber2020body} also used VR to replicate an earlier study on body ownership illusion-based body scaling \citep{van2011being}. While \cite{weber2020body} found significant effects between different body sizes, similarly to our results, they were not able to replicate the difference of judgements between synchronous and asynchronous conditions. However, they did not find any significant difference between the two conditions whereas we found an effect that was inverse to the original study. \cite{maselli2013building} reported that under favorable conditions, even asynchronous stimuli can appear as real to the participants. However, as we did not have any other requirements of the body ownership illusion in place, it seems unlikely that our participants would have somehow experienced the asynchronous condition \change{as more real than} the synchronous condition. In fact, participants often commented on the stimuli appearing to the wrong leg, or "delayed" when we were beginning the application of the asynchronous stimuli. As of now, we are unaware of other VR studies that have attempted to investigate the body-scaling effect using an invisible-body illusion.

Another difference between this study and that of \cite{van2016illusions} is that our invisible body was twice as small \change{as} to the original. One can speculate whether there is a lower-limit to the size of a virtual invisible body that can be inhabited through visuo-tactile stimulus. This lower limit seems unlikely, though, since in their earlier article van der Hoort and Ehrsson \cite{van2011being} demonstrated being able to elicit both the ownership illusion and the perceptual effects using a visible 30 cm body. \change{A limitation of our study compared to that of \cite{van2016illusions} is that we only considered the illusory ownership of a small invisible body instead of having both small and large bodies. It would have been interesting to see if our setup would have caused inverse results regarding the large body condition as well, replicated earlier results, or if some order effects could have appeared.}

One can speculate whether the appearance of the robot manipulating pop tabs somehow caused the inverse effect. However, the robot was hidden most of the time, appearing only during the physics estimation task and disappearing again before the second round of the stimulus and the object size estimation task. In addition, the robot was also accompanied by additional size cues (books, table, pop tabs) so that the participants could clearly see it was doll-sized. Moreover, the robot appeared the same in both synchronous and asynchronous conditions.

It is possible, however, that \textit{something} in the visual appearance of the scene made the difference even if we tried to mimic the layout of the laboratory room presented in the work of \cite{van2016illusions}. Judging from the images presented in the original study, it is difficult to estimate how far the edge of the bed was from the camera, or how exactly it appeared to the participants. It could be that the edge of our virtual exam table was so close to the camera that this (see Fig. \ref{room} top left and right), together with the sensation of laying on the physical exam table, gave participants the sensation of having very short legs from the very beginning of the simulation, even before any stimulus was presented. After this, the visuo-tactile stimulus could have merely scaled up the perceived size of the legs instead of scaling them down as intended. This explanation might be feasible in the sense that \cite{ogawa2017distortion} also reported that environmental cues can also affect perceived size of body cues and not only the reverse, depending on whichever is presented first. In any case, however, \change{further} experiments are necessary to explain the inverse result we found.











\section{Conclusions and future work}
In this paper we investigated the effect of the body ownership illusion on the perception of physics plausibility. We explored the perception of physics plausibility under two conditions that were related to either the presence or lack of \change{a} body ownership illusion caused by synchronous visuo-tactile stimuli. Using the visuo-tactile stimuli, we attempted to elicit the illusion of inhabiting an invisible body roughly the size of a doll (30-40 cm). Similarly to previous research (eg. \cite{van2011being, van2014body, van2016illusions}, we used asynchronous stimuli as a control condition. Our goal was to examine whether correct approximation of physics (\textit{true physics}) or physics that are incorrect and appearing as if the environment is five times larger instead (\textit{movie physics}) appeared more realistic to the participants. We were able to confirm that movie physics did seem to appear more realistic to the participants under the body ownership illusion. However, we were unable to confirm that \textit{true physics} would have appeared more realistic when the body-ownership illusion was lacking. Our exploratory analysis found out that \textit{movie physics} was instead \change{perceived} as real under both conditions.

Moreover, we were not able to replicate the results of \cite{van2016illusions} concerning object size estimations when inhabiting a small invisible body. However, we found a significant opposite effect regarding size estimations; the object sizes were underestimated during the synchronous visuo-tactile condition when compared to the asynchronous condition.

Even though the results were unexpected, they present new information as well as open up avenues for future work regarding the perception of physics, sizes and distances in VR. So far, all our previous studies \citep{pouke2020plausibility, pouke2021plausibility} as well as this one have resulted in \textit{movie physics} appearing as the plausible physics model. This is especially surprising in this study because it included a condition in which neither the scale of the participants nor their perception of sizes and distances, were not manipulated on purpose. We suspect that these results were not due to the existence or lack of a body-ownership illusion, but due the visual appearance and human-like motions of the virtual character. In order to understand the phenomenon regarding plausibility of small-scale rigid body dynamics, it is necessary to conduct more studies involving conditions in which participants are expected, for all intents and purposes, to choose \textit{true physics} as the plausible model. Another surprise was the inverse result regarding object size estimations as the result of body scaling. For this, we suspect that a visual property of the VE, perhaps the length of the virtual exam table, interfered with the participants' sense of \change{their own body size}. An obvious follow-up study to examine the relationship between body-scaling and physics plausibility would be to utilize visible first person avatars of different sizes. Our results show, however, that a different way to represent the physics models is necessary than the one used in this study. In conclusion, further experiments are needed to discover the reasons why unexpected results were obtained regarding both physics plausibility estimations and the inverse effect found in object size estimations. 

\section*{Acknowledgments}
We wish to thank all our participants for volunteering in this study. This work was supported by the Academy of Finland projects PIXIE 331822, PERCEPT 322637, SRC of Academy of Finland project COMBAT 293389, Business Finland project HUMOR 3656/31/2019, and the European Research Council project ILLUSIVE 101020977. \href{https://www.frontiersin.org/articles/10.3389/frvir.2022.869603/full}{\textit{Click here for the published version of this article (Frontiers).}}


\section*{Data Availability Statement}
The raw data supporting the conclusions of this article will be made available by the authors, without undue reservation.

\bibliographystyle{frontiersinSCNS_ENG_HUMS} 
\bibliography{test}

\end{document}